\documentclass[a4paper,11pt]{article}

\usepackage{jinstpub} 
\usepackage{multirow}

\usepackage{lineno}

\usepackage{graphicx}
\usepackage{subfig}
\usepackage{caption}
\usepackage[symbol]{footmisc}

\usepackage{xcolor}

\title{\boldmath The PMT System of the TRIDENT Pathfinder Experiment}

\author[a]{Fuyudi Zhang, \footnote[2]{\label{co_author}These authors contributed equally to this work.}}
\author[c]{Fan Hu,\textsuperscript{\ref{co_author}}}
\author[a]{Shishen Xian,}
\author[a]{Wei Tian,}
\author[d,e]{Kun Jiang,}
\author[a]{Wenlian Li,}
\author[b,a]{Jianglai Liu,}
\author[d,e]{Peng Miao,}
\author[b]{Zhengyang Sun,}
\author[b]{Jiannan Tang,}
\author[d,e]{Zebo Tang,}
\author[b]{Mingxin Wang,}
\author[d,e]{Yan Wang,}
\author[a,b]{Donglian Xu, \textsuperscript{\ref{co_correspondance}}}
\author[a]{Ziping Ye \footnote[1]{\label{co_correspondance} Corresponding authors}}

\affiliation[a]{Tsung-Dao Lee Institute, Shanghai Jiao Tong University, Shanghai 201210, China}
\affiliation[b]{School of Physics and Astronomy, Shanghai Jiao Tong University, Key Laboratory for Particle Astrophysics and Cosmology (MoE), Shanghai Key Laboratory for Particle Physics and Cosmology, Shanghai 200240, China}
\affiliation[c]{Department of Astronomy, School of Physics, Peking University, Bejing 100871, China}
\affiliation[d]{State Key Laboratory of Particle Detection and Electronics, University of Science and Technology of China, Hefei 230026, China}
\affiliation[e]{Department of Modern Physics, University of Science and Technology of China, Hefei 230026, China}

\emailAdd{donglianxu@sjtu.edu.cn, zipingye@sas.upenn.edu}

\abstract{
    Next generation neutrino telescopes are highly anticipated to boost the development of neutrino astronomy. A multi-cubic-kilometer neutrino telescope, TRopIcal DEep-sea Neutrino Telescope (TRIDENT), was proposed to be built in the South China Sea. The detector aims to achieve $\sim0.1$ degree angular resolution for track-like events at energy above 100 TeV by using hybrid digital optical modules, opening new opportunities for neutrino astronomy. In order to measure the water optical properties and marine environment of the proposed TRIDENT site, a pathfinder experiment was conducted, in which a 100-meter-long string consisting of three optical modules was deployed at a depth of $3420~\text{m}$ to perform \textit{in-situ} measurements. The central module emits light by housing LEDs, whereas the other two modules detect light with two independent and complementary systems: the PMT and the camera systems. By counting the number of detected photons and analyzing the photon arrival time distribution, the PMT system can measure the absorption and scattering lengths of sea water, which serve as the basic inputs for designing the neutrino telescope. In this paper, we present the design concept, calibration and performance of the PMT system in the pathfinder experiment.}


\begin{document}
\maketitle
\flushbottom

\section{Introduction}

Neutrino astronomy is entering a golden era, witnessing important breakthroughs in the past decade and yet have more mysteries to be resolved. The first evidence of extraterrestrial high-energy neutrino flux was discovered by IceCube in 2013 \cite{IceCube:2013low}. The identification of TXS 0506+056 in 2017 \cite{IceCube:2018dnn}, a likely high-energy neutrino and cosmic-ray source, was a breakthrough in the young field of multi-messenger astronomy. 
A neutrino hot spot in the diffuse neutrino flux was found in the IceCube 10 years data analysis, correlating with NGC 1068, a nearby active galaxy \cite{IceCube:2022der}. The origin of the high-energy astrophysical neutrinos, the origin of high energy cosmic rays, and the acceleration mechanisms of the cosmic rays, remain unknown \cite{IceCube_selected_results:2022}. Next-generation neutrino telescopes with larger volume and better pointing capability are needed to uncover more astrophysical neutrino sources, reveal the century-old puzzle on high-energy cosmic rays origins and probe the underlining physics.

TRopIcal DEep-sea Neutrino Telescope (TRIDENT) is a proposed next generation neutrino telescope located in the South China Sea \cite{Ye:2022vbk}. It is designed to have an instrumented volume of $\sim 7.5\,\mathrm{km^3}$ and advanced optical modules to achieve first-rate pointing ability and sensitivity. The unique position close to the equator enables detector to scan the entire sky as the Earth rotates. With these advantages, TRIDENT aims to discover more astrophysical neutrino sources and probe fundamental physics with all neutrino flavors. 

In the case of neutrino telescopes with natural water or ice as target material, PMTs are housed in water-resistant glass vessels that are submerged deep beneath water or ice surface for dark environment and cosmic ray shielding. A neutrino telescope comprises of a large number of digital optical modules (DOMs), each containing one or multiple PMTs with readout electronics. The array of PMTs are used to observe the Cherenkov light produced by the secondary charged particles generated in the neutrino-nucleon deep-inelastic scattering \cite{Wiebusch:2003}. Neutrino telescopes, such as IceCube \cite{Hanson:2006bk}, ANTARES \cite{ANTARES:2005hwh} and Baikal-GVD \cite{Avrorin:2016lxq} use one 10-inch PMT within each DOM. 

In recent years, a single glass vessel containing multiple small PMTs is developed and applied in KM3NeT \cite{KM3NeT_mDOM:2014, KM3NeT:2022} and IceCube-Gen2 \cite{IceCube:2022mng, IceCube-Gen2:2021qxw}. Compared to single-PMT DOMs, multi-PMT DOMs can receive photons from all directions, potentially providing better pointing capability. Multi-PMT DOMs can also allow local coincidence trigger to reduce background rate \cite{Trovato:2014msl}. TRIDENT will adopt an innovative design called hybrid Digital Optical Module (hDOM), which uses both multiple small PMTs and arrays of silicon photomultiplier (SiPMs) to increase the photon detection area and gain faster time response \cite{Hu:2021jjt}.

The TRIDENT pathfinder experiment (TRIDENT EXplorer, T-REX for short) aims to measure the optical properties of the deep-sea water and investigate the marine environment at the selected site, which will serve as a basis for designing the proposed neutrino telescope. A PMT detection system is designed for T-REX to measure the water optical properties and the level of backgrounds. This paper presents the selection, testing, and calibration of the PMTs used in T-REX \cite{Ye:2022vbk}, and reports the performance of the PMT detection system in the sea experiment. The remaining parts of this paper are structured as follows: The basic idea and design of T-REX is introduced in section \ref{sec:pathfinder}. Section \ref{sec:optical_module} describes the inner structure and assembly of the light receiver module in T-REX. Tests on a sample of PMTs were conducted and the results are presented in section \ref{sec:pmt_selection}. Section \ref{sec:calibration} shows the calibration results of the selected PMTs in a temperature-controlled dark room. Finally, the performance of the PMTs in the pathfinder experiment is presented in section \ref{sec:performance_pathfinder}. Summary and outlook are given in section \ref{sec:conclusion}.

\section{The TRIDENT Pathfinder Experiment} \label{sec:pathfinder}

\begin{figure}[ht!]
    \centering
    \includegraphics[width=6cm,height=10.5cm]{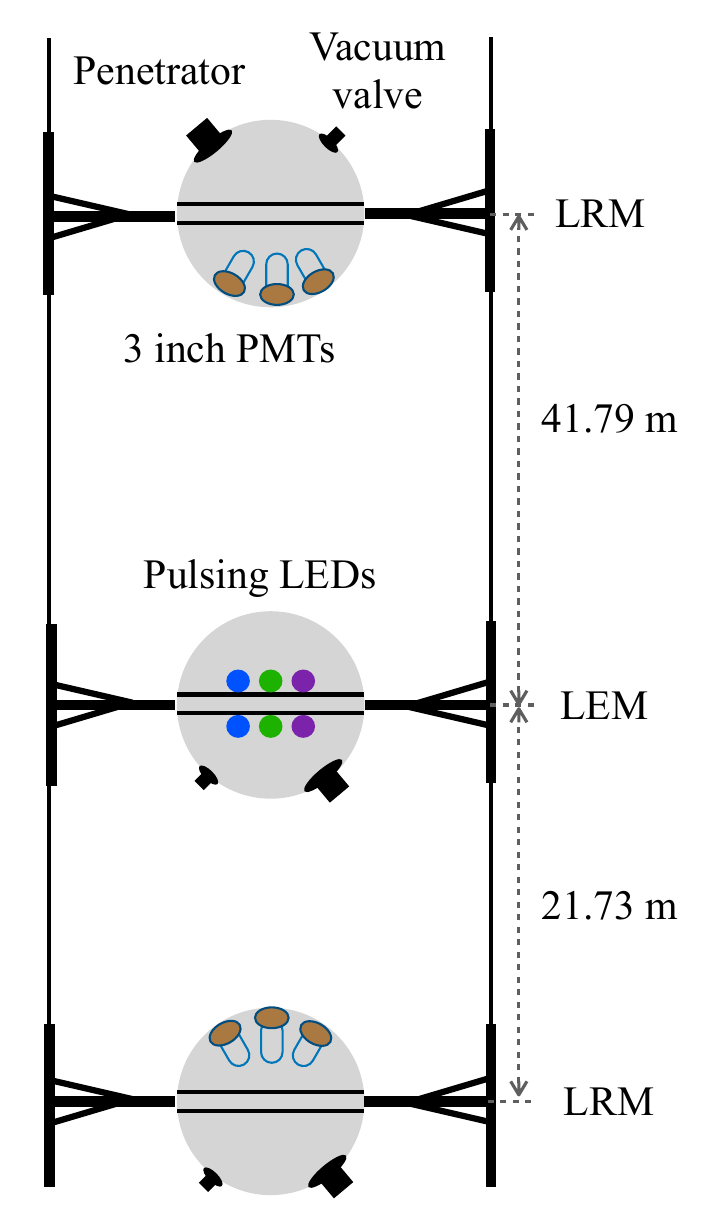}
    \caption{The detection apparatus of T-REX. The glass vessel in the middle is the light emitter module (LEM). It is equipped with LEDs and double-layer diffuser that makes it an isotropic light source. Two light receiver modules (LRMs) each housing three 3-inch PMTs and one camera are placed at different distances away from the LEM to conduct a near-far relative measurement.}
    \label{fig:detection_unit}
\end{figure}

The optical properties of the water and the ambient light background are important for the design of a neutrino telescope, which relies on the simulation of Cherenkov light propagation in water and the reconstruction of neutrino interaction using the photo-sensor signals. The T-REX was conducted at the pre-selected site in the South China Sea to measure the optical properties of deep-sea water \textit{in-situ}, as well as the ambient light background produced predominantly by $^{40}\text{K}$ decay and bioluminescence in seawater. Photons propagating in water might undergo absorption and/or scattering processes. Absorption will reduce the number of photons arrive on PMTs, while scattering can alter the arrival time and direction of the photons. T-REX will perform a near-far relative measurement to determine the absorption length, and extract the scattering length from the photon arrival time distribution observed by PMTs. The level of ambient light background due to $^{40}\text{K}$ radioactivity and bioluminescence can be measured by using correlated signals on multiple PMTs \cite{KM3NeT:2014wix, Bailly:2021mei}. 

As shown in Figure \ref{fig:detection_unit}, the T-REX apparatus consists of three 17-inch spherical glass vessels mounted on a rigid frame and connected with optic-electric cables for power supply and data transmission. 
These vessels are water-tight and able to withstand high pressure, protecting PMTs and electronics inside. The glass vessel at the middle is a light emitter module (LEM), equipped with light-emitting diodes (LEDs) at various wavelengths and a double-layer diffuser that makes it an isotropic light source \cite{Li:2023wqk}. Two identical LRMs are housed within the two glass vessels positioned at different distances from the LEM, one being $41.79\,\text{m}$ above and the other $21.73\,\text{m}$ below. Each LRM consists of three 3-inch PMTs, one camera, and readout electronics.

For the PMT detection system, the corresponding LEDs in the LEM operate in a pulsing mode. The LED pulsing frequency is 10 kHz. Each light pulse has a width of $\sim 3 \,\mathrm{ns}$ and emits $\sim10^9$ photons \cite{Tang:2023jmn, Li:2023wqk}. The PMTs in the LRMs are externally triggered by a White Rabbit (WR) system in synchrony with the pulsing LEDs at the same frequency, so that the PMTs can detect the photons emitted by the LEDs. The two LRMs are at different distances away from the LEM, allowing us to perform relative measurements of the optical properties of the water \cite{ANTARES_measurement:2004, P-One_measurement:2021}.

\section{Light Receiver Module} \label{sec:optical_module}

\begin{figure}[htb]
    \begin{minipage}[t]{.36\textwidth}
        \centering
        \includegraphics[width=\textwidth]{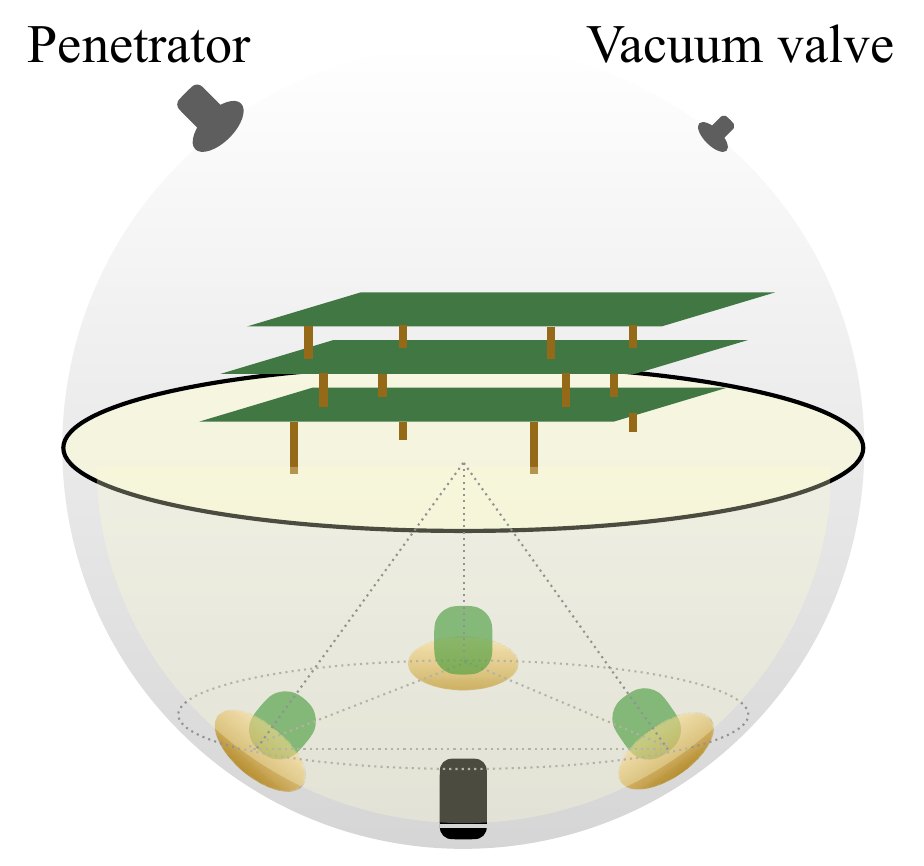}
    \end{minipage}
    \hfill
    \begin{minipage}[t]{.38\textwidth}
        \centering
        \includegraphics[width=\textwidth]{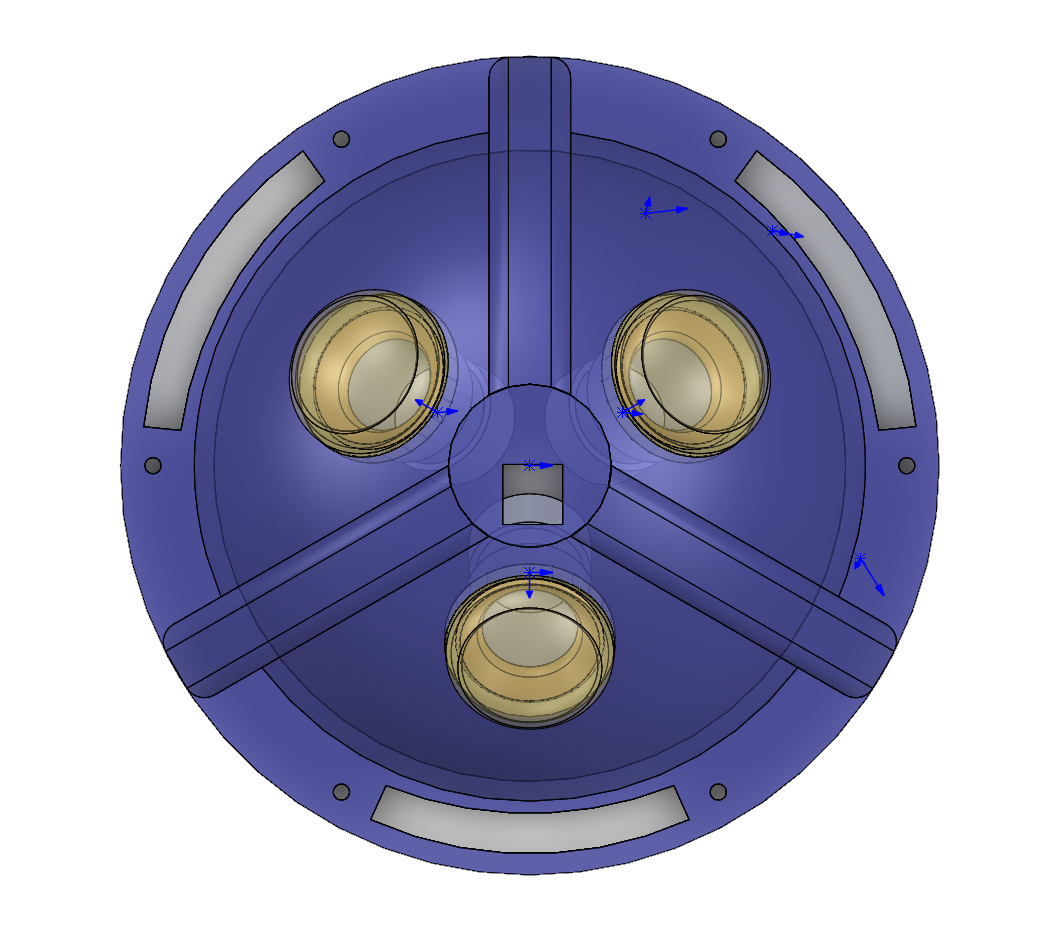}
    \end{minipage}  
    \caption{\textbf{Left}: A sketch of a light receiver module (LRM). Optical sensors are housed in the lower hemisphere and glued in optical gel, while electronic boards are installed in the upper hemisphere. \textbf{Right}: 3D-printed support structure for PMTs and camera.}
    \label{fig:receiver_module}
\end{figure}

Each of the two identical LRMs contains three 3-inch PMTs to observe the light emitted by the pulsing LEDs and one camera to observe the steady LEDs in the LEM. The PMT system and camera system work independently, the camera and steady LED system will be turned off while the PMT and pulsing LED system is working. This paper focuses on the PMT system, for details of the camera system please refer to \cite{TianWei:2022}. In each LRM, the PMTs are installed on a support structure in one hemisphere, which will face the LEM for observation. A drawing of the LRM is shown in the left panel of Figure \ref{fig:receiver_module}, and the 3D-printed support structure is shown in the right panel. The three PMTs are oriented at an angle of 30$^{\circ}$ from the vertical central axis of the LRM, and they are positioned uniformly with 120$^{\circ}$ separation. The space between the glass sphere and the PMTs is filled with highly transparent silicone rubber gel to ensure smooth propagation of photons between different media before reaching the PMT photocathode.

The other hemisphere of the glass vessel hosts the electronic boards for managing power supply, pre-amplifier, analog to digital coverter (ADC), network nodes for data transmission and slow control unit \cite{Wang:2023rvb}. A penetrator and vacuum port are also installed in this hemisphere. The PMT working voltage is set to $\sim1375\,\text{V}$, resulting in a gain of $\sim10^7$. The PMT signals are amplified by 10.5 times with the pre-amplifier and then digitized by the ADC at a sampling rate of $250~\text{MHz}$.
The digitized data is transmitted to the Data Acquisition (DAQ) computer on the scientific research vessel through optical fibers in an armoured cable that connects the T-REX apparatus to the ship. Clock synchronization among the three modules (one LEM and two LRMs) is achieved by the WR system.

The LEM is up-down symmetric, with three pulsing LEDs at wavelengths of $405\,\text{nm}$, $450\,\text{nm}$ and $525\,\text{nm}$ in each hemisphere to trigger the PMTs. A 3D-printed photosensitive resin spherical shell encompasses all the LEDs, making the emission sphere an isotropic light source. Small difference of the two hemispheres of the LEM exists due to manufacturing quality and mechanical components, including penetrator and vacuum port. The resulting brightness ratio of the two hemispheres is calibrated in the lab, which will be inputs to the data analysis of the near-far relative measurement \cite{Li:2023wqk}.

\section{PMT Pre-Selection} \label{sec:pmt_selection}

For successful measurement of the deep-sea water optical properties, the PMTs are required to be able to count the number of detected photons and measure their arrival times with good precision, as well as providing high signal-to-noise ratio. The PMTs are thus selected based on the overall performance that includes peak to valley ratio (PVR), transit time spread (TTS), quantum efficiency (QE), dark count rate (DCR) and single photoelectron (SPE) voltage. We choose the XP72B22 PMT manufactured by Hainan Zhanchuang Photonics Technology Co., Ltd (HZC) for T-REX. Specifically, this PMT is custom designed for JUNO with a glass bulb shape that optimizes light collection efficiency and TTS \cite{Cao:2021wrq}. It has a low dark count rate of $\sim 1~\text{kHz}$ and excellent SPE resolution of $\sim 30\%$.

\begin{figure}[htb]
    \centering
    \includegraphics[width=0.80\textwidth, height=180mm]{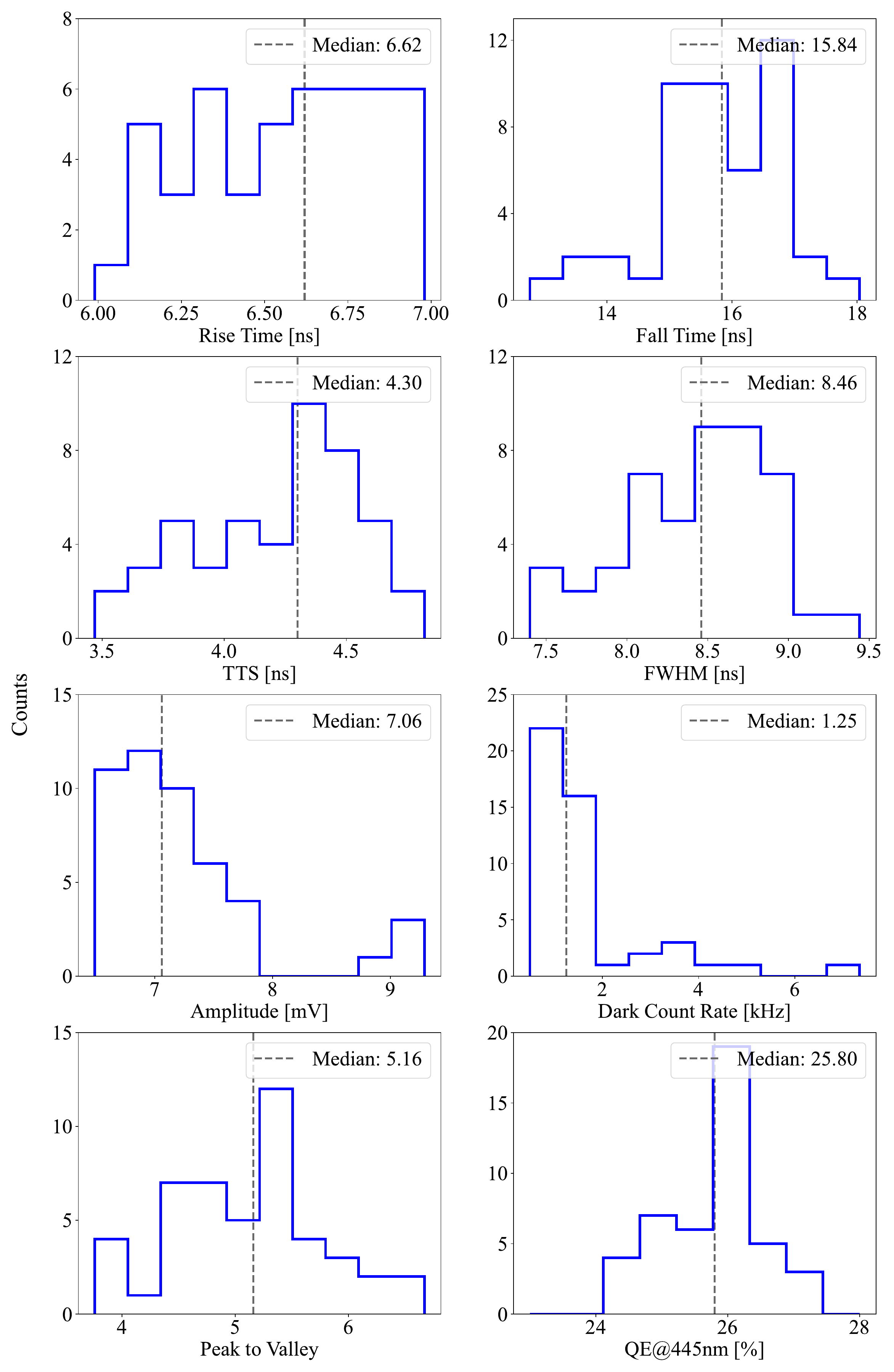}
    \caption{Measurement results of a sample of 50 PMTs. Properties, including rise time, fall time, transit time spread (TTS), full width half maximum (FWHM), single photoelectron (SPE) amplitude, dark count rate (DCR), peak to valley ratio (PVR), and quantum efficiency (QE) at 445 nm are measured. These properties are used to select PMTs with the best overall performance for T-REX. }
    \label{fig:ustc_pmt}
\end{figure}

\begin{figure}[htb]
    \centering
    \includegraphics[width=0.70\textwidth]{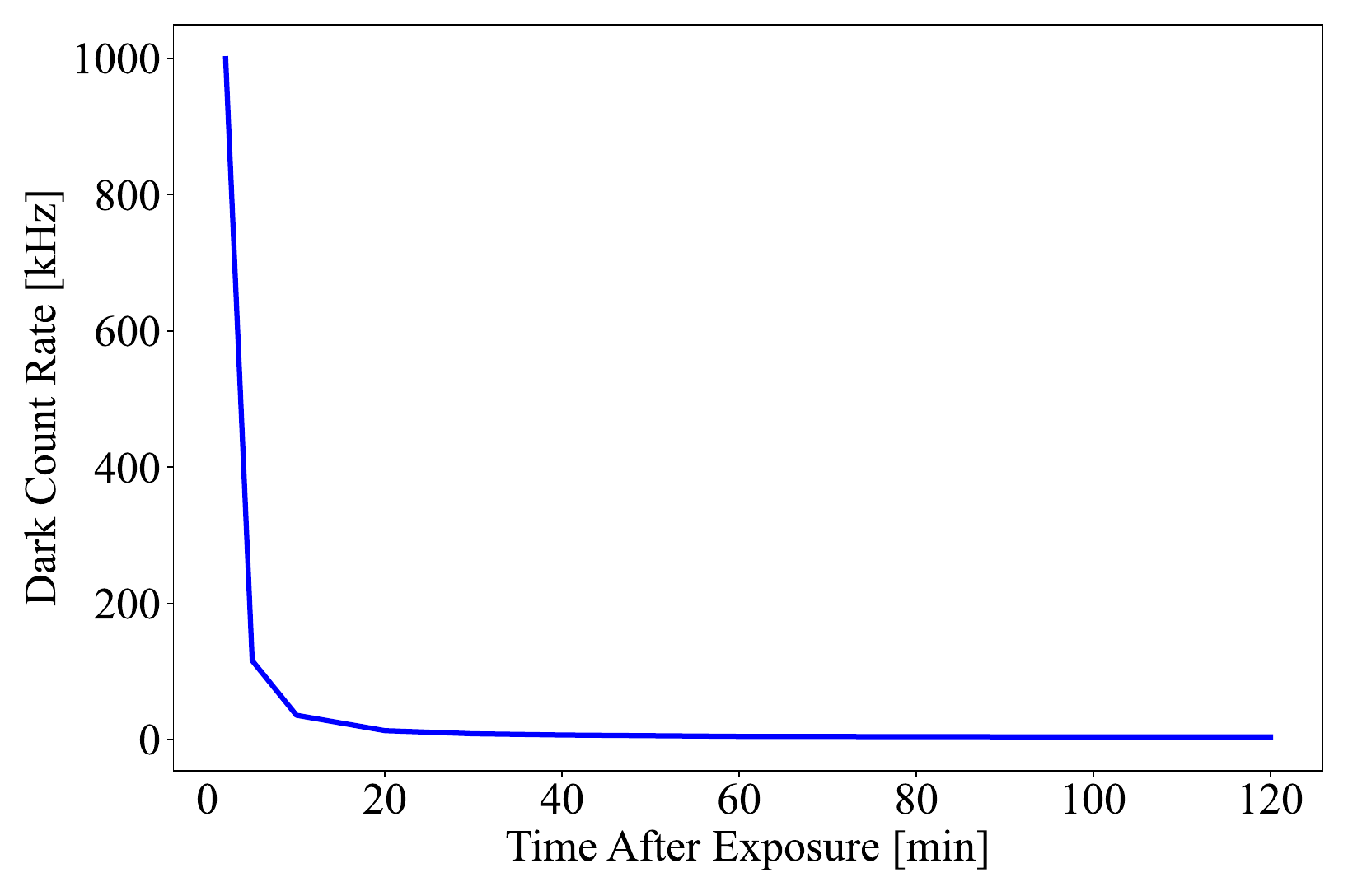}
    \caption{Measured dark count rate as a function of time after light exposure.}
    \label{fig:exposure_time} 
\end{figure}

We selected six PMTs to be installed in the two LRMs, based on the testing results of 50 PMTs, using a PMT test facility designed for LHAASO \cite{Jiang:2020rdv}. Each PMT is exposed to a pico-second-width laser pulse through an optical fiber. The LabView program controls the test system, which first determines the gain-voltage relation for each PMT. The photocathode characteristics, including rise time, fall time, TTS, SPE voltage, DCR, PVR and QE are then measured at different gains. The statistics of the testing results of the 50 PMTs measured at a gain of $\sim 10^7$ is summarized in Figure \ref{fig:ustc_pmt}. We then selected six PMTs that have the best overall performance, and it turned out that the selected PMTs have QE > 25$\%$ at $445~\text{nm}$, DCR $< 1.25~\text{kHz}$, TTS $< 4.0~\text{ns}$, PVR $> 5.0$ and SPE voltage $> 6.0~\text{mV}$, which satisfy the requirements for T-REX.
Since the PMTs will inevitably be exposed to light during the deployment process, we need to confirm the PMT cooling down times are short enough for deployment process in the sea experiment. Measurement of the PMT cooling down time was done in the lab, where each PMT was exposed to the same amount of light with an illuminance of 15 lux for $15~\text{minutes}$, the DCR as a function of time after the exposure was measured. A typical cooling down curve is shown in Figure \ref{fig:exposure_time}, showing that the PMTs' cooling down times are $\sim$ 30 minutes, much shorter than the deployment time. So the PMT system can begin the measurement in a state with steadily low dark noise rate once it reaches the experiment depth. 

\section{PMT System Calibration at Low Temperature} \label{sec:calibration}

\begin{figure}[ht] 
    \centering
    \includegraphics[width=0.99\textwidth]{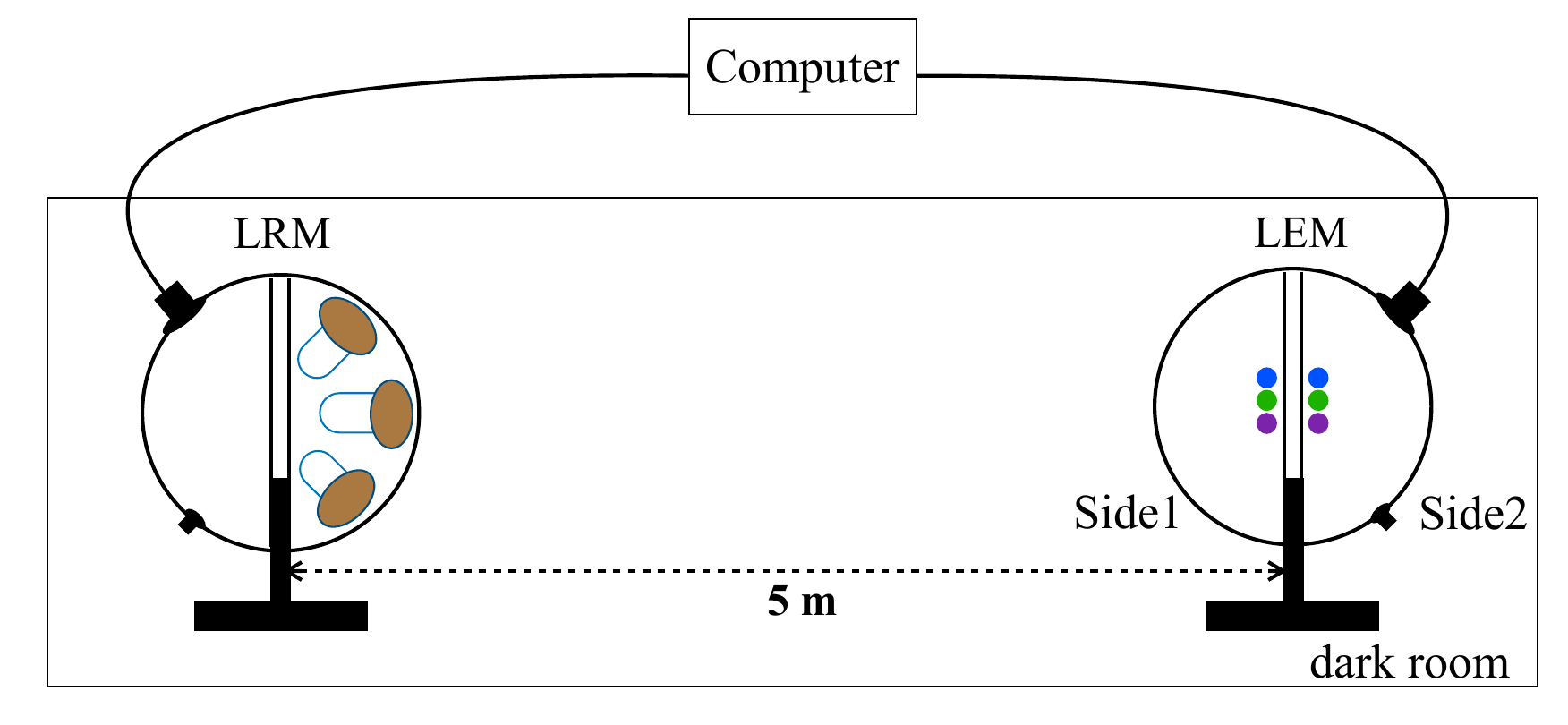}
    \caption{Experimental setup in the temperature-controlled dark room for the PMT system calibration.}
    \label{fig:exp_setup}
\end{figure}

The temperature of sea water is depth-dependent and it is about $2^{\circ}\text{C}$ at depths below 2800 m \cite{Ye:2022vbk}. Since PMTs' and LEDs' performance are temperature dependent, we calibrated their properties in a dark room at about $2^{\circ}\text{C}$ after they had been installed in the LEM and LRMs. In the calibration experiment, we measured the relative photon detection efficiency (PDE) of PMTs, the brightness ratio of LEDs on the two sides of LEM, and the photon arrival time distribution (which captures the intrinsic time response of LED and PMT) in air. These calibrated quantities are critical inputs to the analysis of T-REX sea experiment data for measuring the water optical properties. 

The calibration experiment is conducted in a 10-meter long temperature-controlled dark room, where the LEM and LRM are placed $5.0~\text{m}$ away from each other, as illustrated in Figure \ref{fig:exp_setup}. To measure the brightness ratio of the two hemispheres of LEM, each of them acts as an independent light source, labelled as Side1 and Side2, to trigger the same PMT. To measure the relative PDE of PMTs, two LRMs in turn observe the light emitted from the same side of the LEM. The six PMTs are labelled with IDs 1,2,3 and 4,5,6 in the far and near LRM, respectively. In total, we set up four different combinations of emitter and receiver, numbered as GroupID = 1, 2, 3 and 4  in Table \ref{tab:exp_config}. Specifically, GroupID 1 and 4 are the combinations used in T-REX sea experiment. To counteract the residual asymmetric effect of the three PMTs' orientations in the LRM, the measurement for each group is repeated three times by rotating the LRM along their central axis at three different angles (0$^{\circ}$, 120$^{\circ}$, 240$^{\circ}$). This yields a total of twelve measurement groups. The results presented in the following are the averaged results of the three rotation angles.

\begin{table}[htb]
    \centering
    \begin{tabular}{|c|c|c|c|}
        \hline
        GroupID & Emitter  & Receiver & PMT IDs\\
        \hline                            
        1  & Side1   & far LRM & 1,2,3       \\
        2  & Side2 & far LRM & 1,2,3   \\
        3  & Side1   & near LRM & 4,5,6     \\
        4  & Side2 & near LRM & 4,5,6    \\
        \hline
    \end{tabular}
    \caption{Measurement groups in the calibration experiment. Two hemispheres of the LEM are labelled as Side1 and Side2, serving as independent light sources in the calibration experiment. Two LRMs are labelled with near and far to indicate their distances from the LEM in T-REX sea experiment. The GroupID 1 and 4 are the combinations used in sea experiment.}
    \label{tab:exp_config}
\end{table}

\begin{figure}[htb]
    \centering
    \includegraphics[width= 0.82\textwidth]{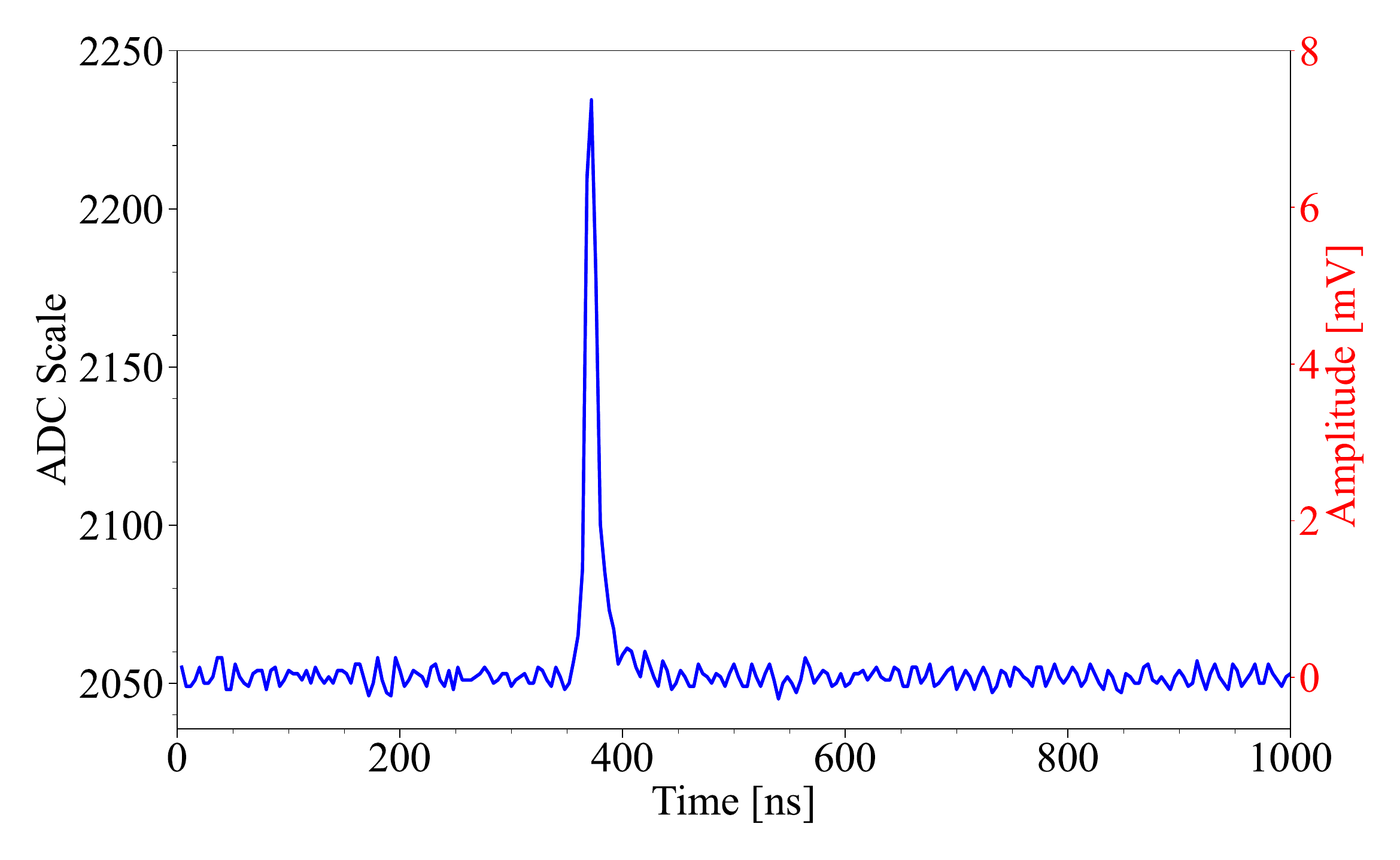}
    \caption{
    A PMT waveform digitized by an ADC with a sampling rate of $250~\mathrm{MHz}$.
    The y-axis on the left and right indicates the same waveform expressed in ADC scale and in unit of millivolt, respectively.
    }
    \label{fig:waveform}
\end{figure}

In the calibration experiment, the DAQ system is configured to match that of T-REX sea experiment, with the three PMTs in each LRM and pulsing LEDs in the LEM triggered externally at a rate of 10 kHz. The PMT signals are digitized by a $250~\mathrm{MHz}$ ADC, resulting in a $1000~\mathrm{ns}$ DAQ window. Figure \ref{fig:waveform} shows an example of the PMT waveform. The procedure of analyzing PMT waveforms in calibration experiment is the same as that in sea experiment. 

\subsection{Single Photoelectron Amplitude and Charge Measurements}
 
\begin{figure}[htb]
    \centering
    \includegraphics[width= 0.80\textwidth]{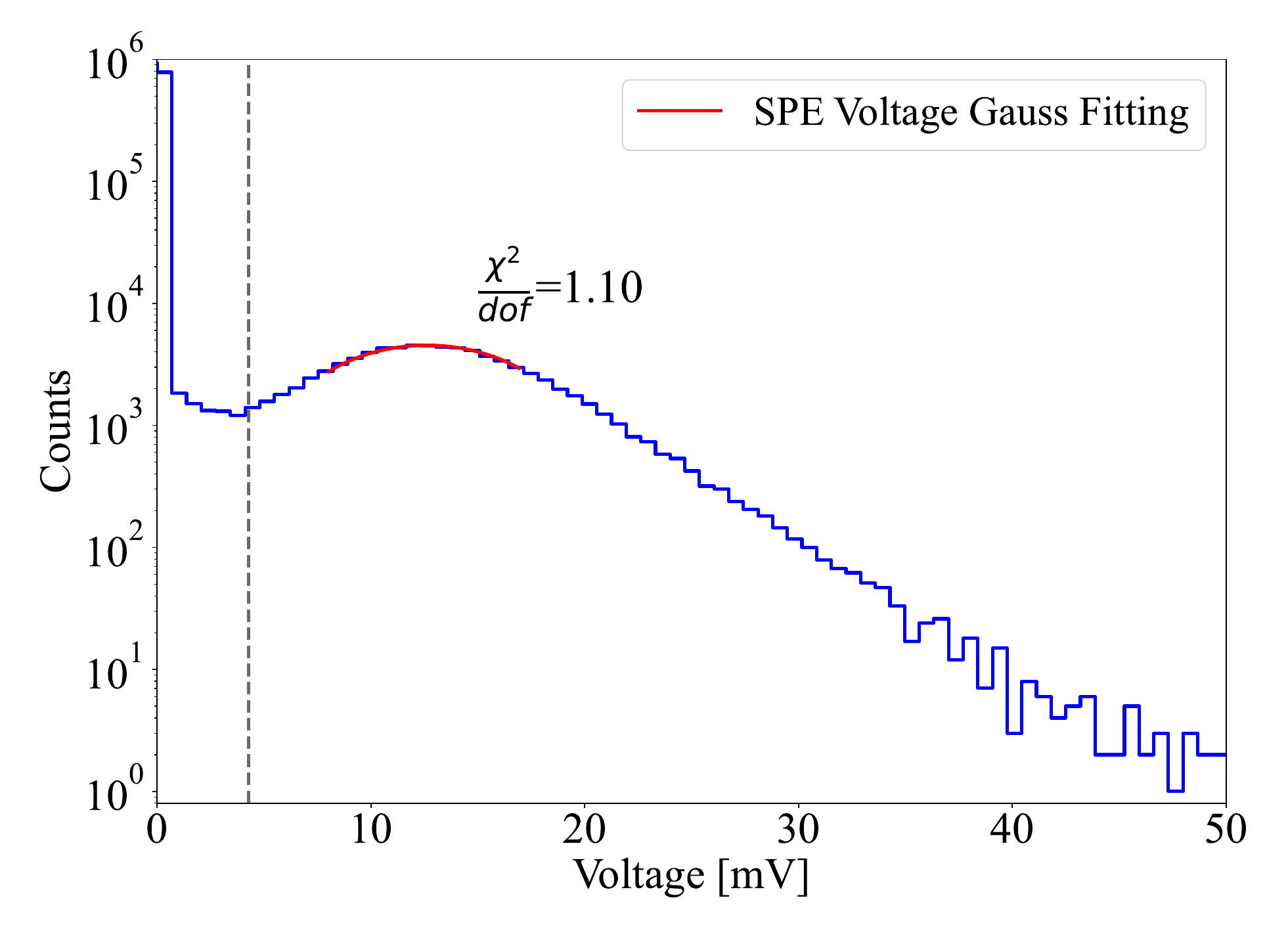}
    \caption{The peak voltage distribution of PMT waveforms. The voltage distribution is fitted with a Gaussian function and the SPE amplitude is determined as the mean value of the Gaussian function. The vertical dashed line at 1/3 of SPE amplitude shows the threshold for pulse triggering.}
    \label{fig:spe_amplitude}
\end{figure}

Understanding the SPE response of the PMT is important for triggering and counting the number of detected photons. The SPE amplitude is the average voltage of SPE signals. We first find the maximum voltage of the waveform in each DAQ window by: 

\begin{equation}
    V_{\text{max}} = (A_{\text{max}} - A_\text{baseline}) \cdot \frac{\delta V}{a},
\end{equation}

\noindent where $A_{\text{max}}$ and $A_\text{baseline}$ are the maximal ADC point and ADC baseline in each DAQ window, $\delta V = 0.43\,\mathrm{mV}$ is the conversion factor from ADC scale to voltage, $a=10.5$ is the pre-amplifier factor. By stacking analysis of a large number of waveforms, we get the distribution of the maximum voltage, as shown in Figure \ref{fig:spe_amplitude}. The SPE amplitude is obtained by fitting the voltage distribution with a Gaussian function. The dashed vertical line at 1/3 of SPE voltage shows the threshold for triggering PMT signal pulses.

\begin{figure}[htb]
    \centering
    \includegraphics[width=0.84\textwidth]{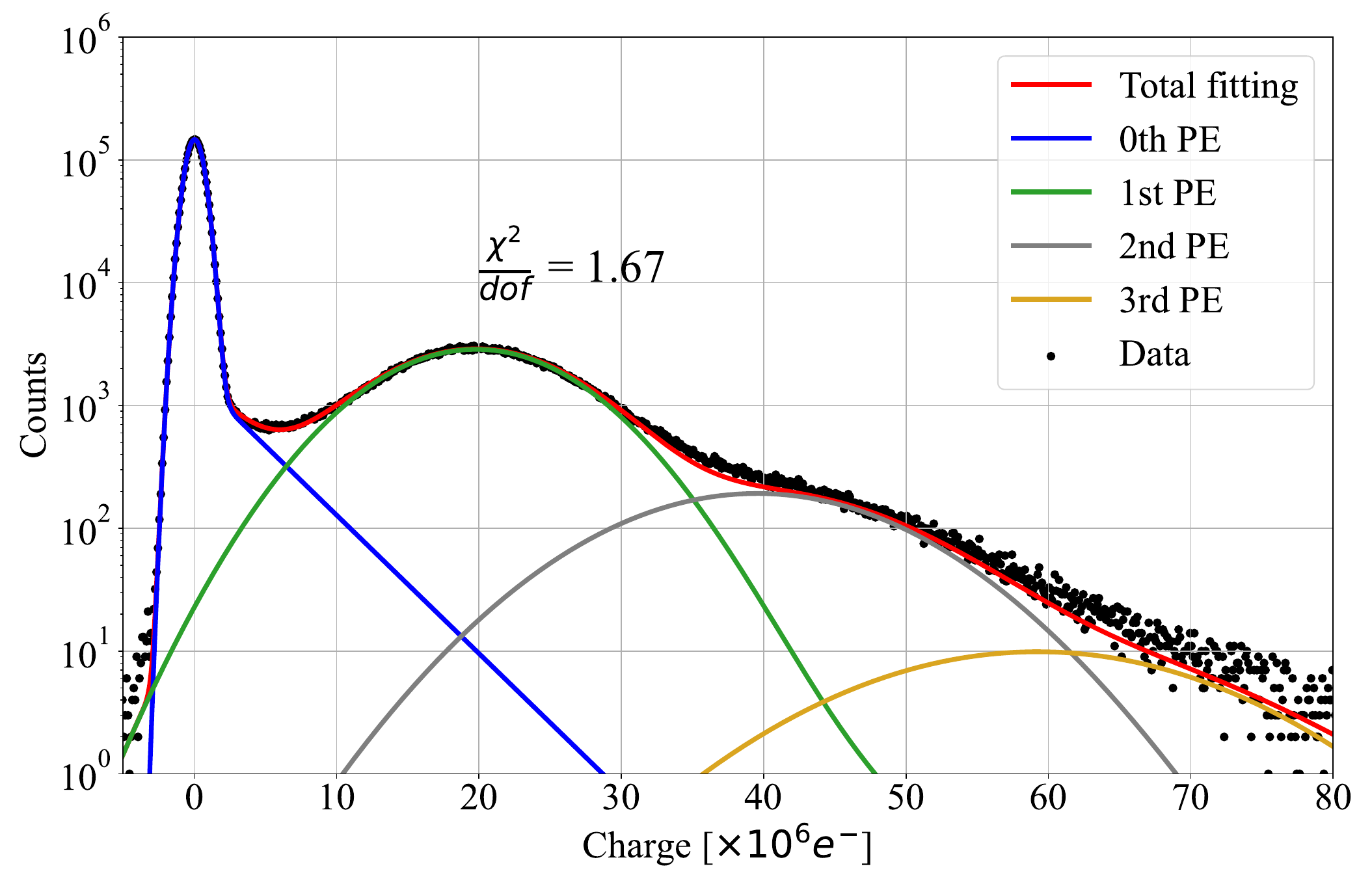}
    \caption{An example of charge distribution of PMT$\#4$ obtained in the calibration experiment. The charge distribution is well-fitted with the function up to 3PE.}
    \label{fig:charge_data}
\end{figure}

The SPE charge (PMT gain) can be obtained by fitting the charge distribution of a PMT. The charge integral of a waveform is calculated by:
\begin{equation}
    G = \sum_{i}\frac{V_i \cdot \delta t}{R \cdot e}
\end{equation}
\noindent Here, $V_i$ is the voltage at the $i$-th time bin; $R=50~\Omega$ represents the ADC load resistor, i.e. the oscilloscope input impedance; $\delta t=4\,\text{ns}$ is the sampling time step in nanoseconds; $e$ is the charge of an electron. An example of the charge distribution is shown in Figure \ref{fig:charge_data}, which is fitted with the following model \cite{Bellamy:1994bv}: 

\begin{equation}
    \begin{aligned}
        f(Q) = & \sum_{n=0}^{\infty} \text{Poisson}( n , \mu ) \times \\
        & \left( (1-w) \times \frac{1}{\sqrt{2\pi} \sigma_n} e^{-\frac{(Q-Q_n)^2}{2\sigma_n^2}} + 
        w \times  \frac{\lambda}{2} e^{ \frac{ ( \lambda \sigma_n )^2 }{2} } e^{ - \lambda ( Q - Q_n ) } \cdot \text{Erfc} \left( \frac{1}{\sqrt{2}} \left( \lambda \sigma_n - \frac{ Q - Q_n }{ \sigma_n } \right) \right) \right) ,
    \end{aligned}
\end{equation}

\noindent where a Poisson distribution with mean value $\mu$ is assumed for the number of photoelectrons (nPEs) observed by the PMT.
The second line of the equation describes the charge distribution function for nPEs. It is composed of a Gaussian function and an exponentially modified Gaussian distribution. 
$Q_{n} = Q_{0} + n \cdot Q_{1} $ and $\sigma_{n} = \sqrt{ \sigma_{0}^{2} + n \cdot \sigma_{1}^{2} }$ are the mean value and the standard deviation of Gaussian function for nPEs, respectively. 
$w$ is the probability of dark current background present in the fixed window. The exponentially modified Gaussian distribution is the convolution of an exponential function with a Gaussian function, where $\lambda $ is the exponential decay parameter and $\text{Erfc}(x) $ is the complementary error function. It is used to represent the exponential distribution of partially amplified photoelectrons. It can be seen that the charge distribution is well fitted with the model and the PMT gain is determined as $Q_{1}$. Once the PMT gain is obtained, we will be able to count the number of detected photons for each PMT waveform, which will serve as the basis for calibrating the relative PDE of PMTs and the brightness ratio of LEDs.

\subsection{Photon Detection Efficiency}
\label{subsec:calibrated_pde}

In the sea experiment, the attenuation length of sea water is derived from measuring the ratio of the averaged number of photons detected by a far PMT and a near PMT. For the relative measurement, the LED brightness ratio and the PMT relative detection efficiency must be subtracted. After deducting the apparatus' effect (the PMTs' detection efficiency and the intensity of the LED pulsing), the optical properties of water can be extracted. The LED brightness ratio and PMT relative detection efficiency are measured after assembling the LEM and LRM, so that the systematic uncertainties due to reflection and absorption effects of intermediate surfaces are also included in the calibration. 

\begin{figure}[ht] 
    \centering
    \includegraphics[width = 0.94\textwidth]{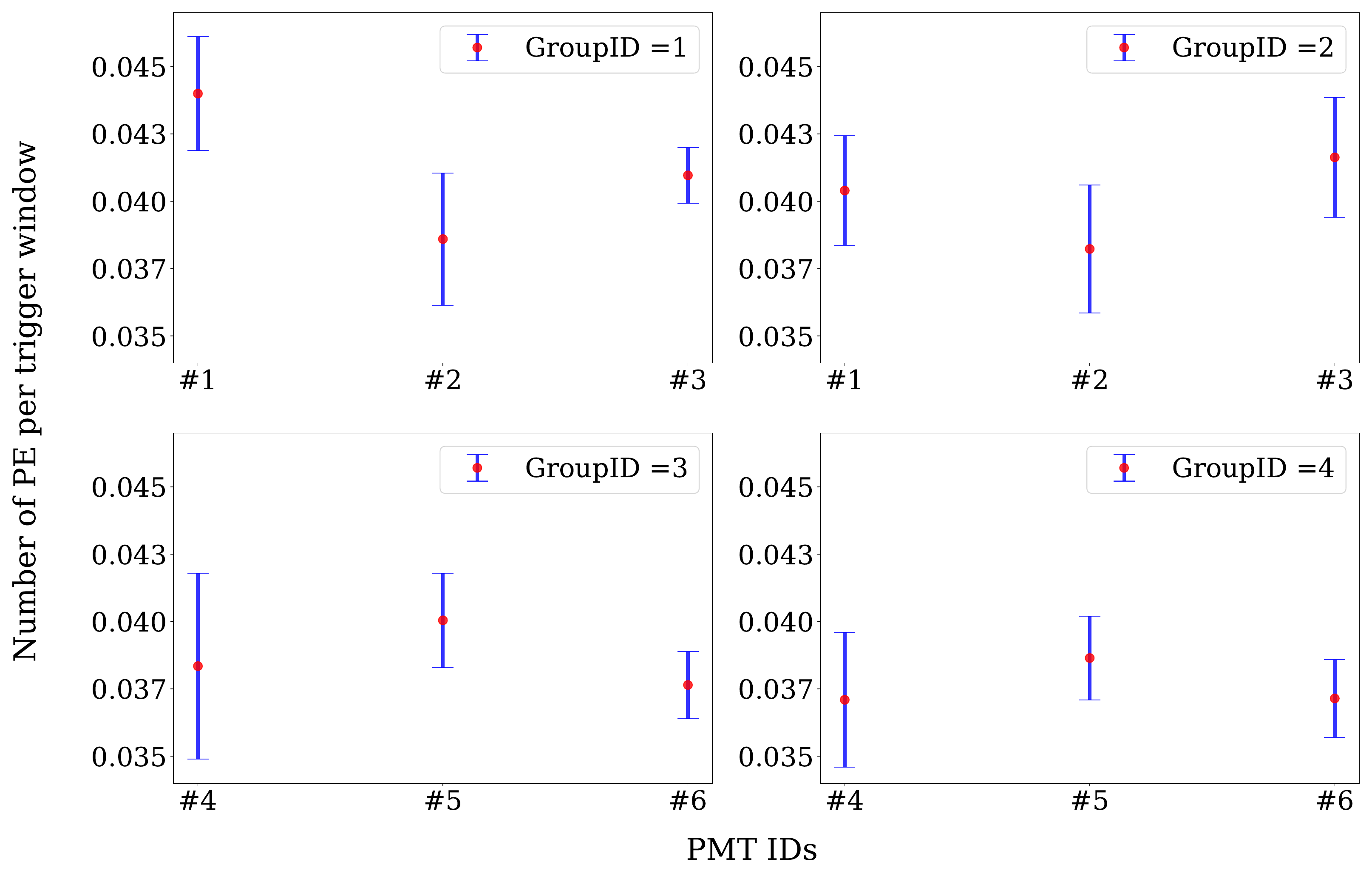}
    \caption{Averaged number of photoelectrons (PEs) per trigger window for different measurement groups listed in Table \ref{tab:exp_config}.}
    \label{fig:photon_number}
\end{figure}

We obtain the number of photoelectrons (PEs) for each PMT waveform, using the following analysis procedure. A fixed window is defined for LED photon signal region, and the charge integral, $Q_\text{p}$, is calculated for the region. The number of PEs is determined by $N_{PE} = Q_\text{p}/G$, where $G$ is the PMT gain. By stacking analysis of a large number of waveforms for each PMT, the averaged number of PEs observed by the PMT can be obtained, which is proportional to the product of the LED brightness and the PMT detection efficiency.

\begin{figure}[ht]
    \centering
    \includegraphics[width = 0.76\textwidth]{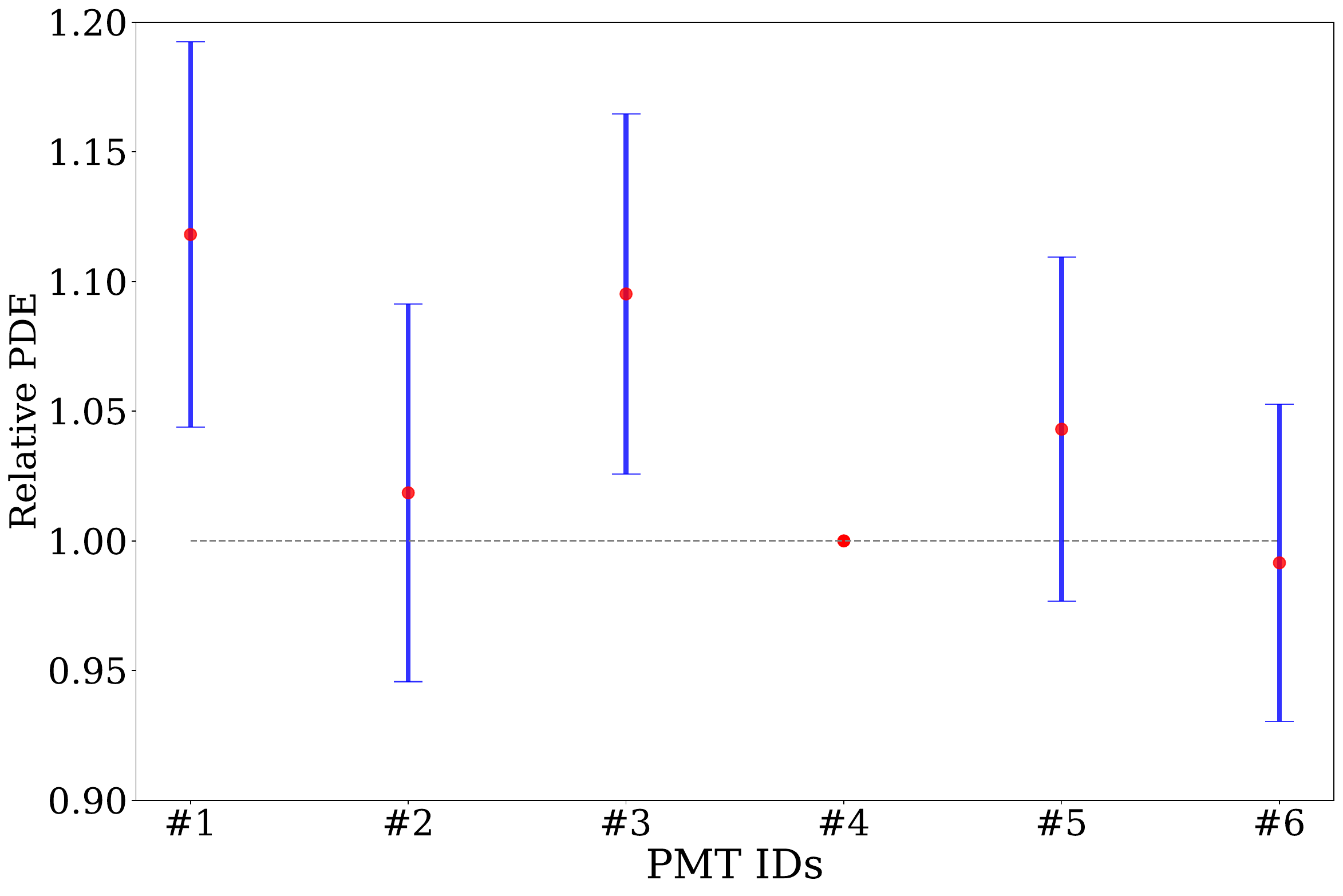}
    \caption{Relative photon detection efficiencies (PDEs) of the other five PMTs compared to PMT$\#4$. The efficiencies shown here are the averaged results of using the two sides of the LEM as independent light sources.}
    \label{fig:detection_eff}
\end{figure}

The average number of PEs per trigger window detected by each PMT for each measurement group is shown in Figure \ref{fig:photon_number}. 
The brightness ratio of the two sides of LEM is calculated by comparing the number of PEs detected by the same PMT for the two sides of LEM, namely comparing GroupID 1 to GroupID 2, and GroupID 3 to GroupID 4). The measured brightness ratio at the wavelength of $450~\mathrm{nm}$ is about $1.028 \pm 0.041$ \cite{Li:2023wqk}. 
The relative PDE of the PMTs is calculated by comparing the number of PEs of two PMTs triggered by the same light source. 
PMT$\#4$ is chosen as the baseline, and the relative PDEs of the other PMTs with respect to PMT$\#4$ are shown in Figure \ref{fig:detection_eff}.

\subsection{Photon Arrival Time Distribution in Air}
\label{subsec:time_distribution}

The measured photon arrival time distribution in sea water is a convolution of the detector's intrinsic time response and the scattered photon time profile. The scattering effect in sea water would result in a longer tail in the distribution. The detector's intrinsic time response is identical to the photon arrival time distribution measured in lab, where the scattering effect is negligible in air. To extract the scattering photon time profile from the observed distribution in sea water, we first need to measure the detector's intrinsic time response, which includes the LEM emission and LRM detection time response.

\begin{figure}[htb]
    \centering
    \includegraphics[width=0.86\textwidth]{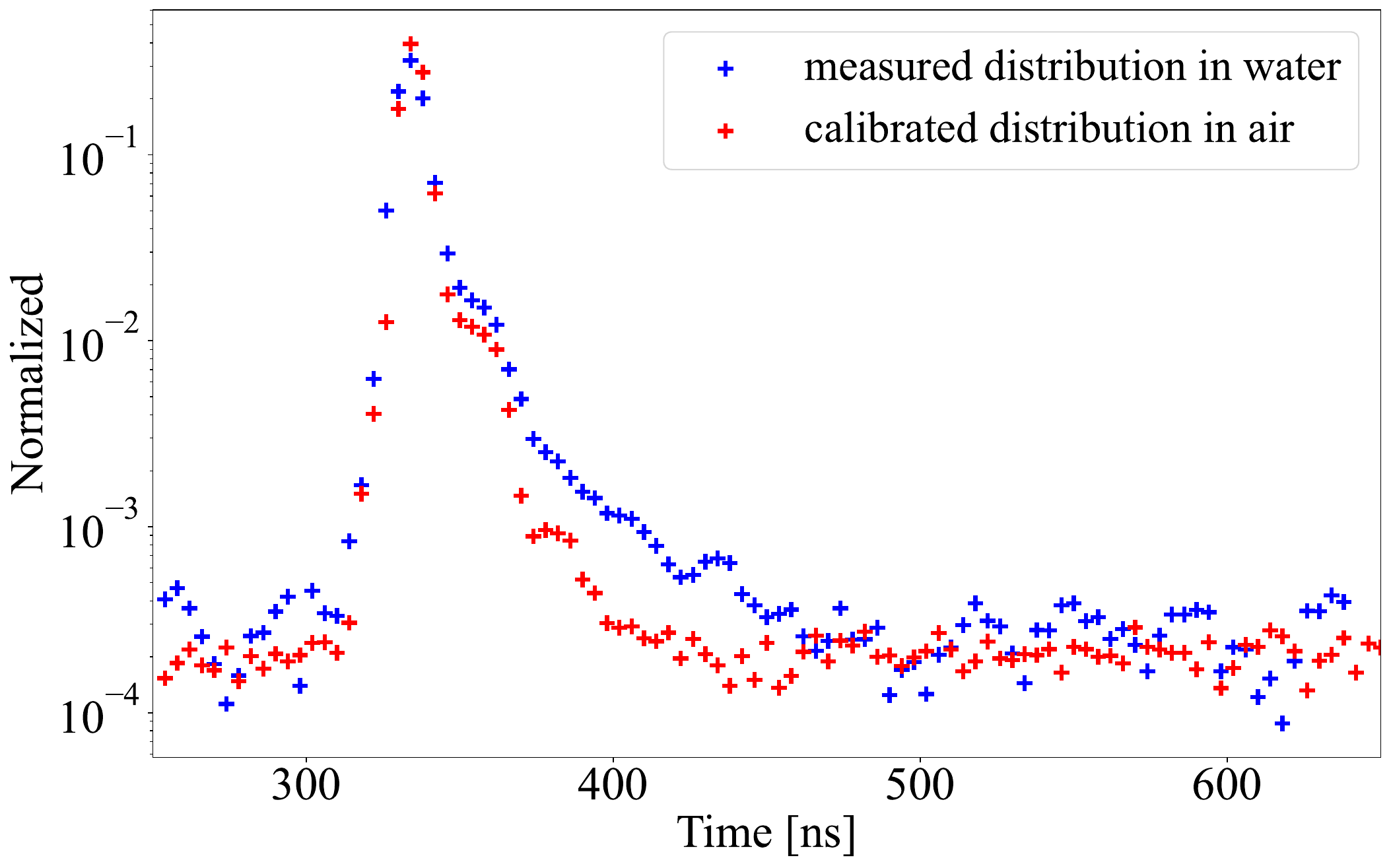}
    \caption{The comparison of photon arrival time distributions in water with T-REX (blue) and the calibrated distribution in air (red) measured by the same PMT. The two distributions are normalized. The longer tail of the distribution in deep-sea water shows scattering effect.}
    \label{fig:arr_time}
\end{figure}

The photon arrival time distribution is obtained by stacking the arrival time of LED signal pulses for a large number of DAQ windows. To extract the distribution, multi-level data processing on digitized waveforms are conducted, which is the same as the analysis procedure applied to the sea experiment data, keeping a consistent workflow and avoiding potential bias in data analysis, for details of the analysis procedure please refer to \cite{Ye:2022vbk}. Figure \ref{fig:arr_time} shows the photon arrival time distributions measured in deep-sea water with T-REX and the calibrated one in air for the same PMT. Both distributions are normalized and the calibrated distribution in air is shifted so that its raising edge is aligned with the sea experiment distribution. 
The sharp peak observed at around $350\,\mathrm{ns}$ corresponds to the arrival time of LED emitted photons at PMT, and its intrinsic spread is caused by the LED emission and PMT detection time response. The shoulder-like structure at $\sim30\,\mathrm{ns}$ after the main peak is due to the after-pulsing effect of the PMT. The longer tail of the distribution in sea experiment shows the scattering effect of photons in seawater. The photon arrival time distribution in air is adopted as the probability density function (PDF) of the LEM and LRM time response. The photon propagation time spread PDF in water, which describes the distribution of photon arrival time due to absorption and scattering of photons in water and will depend on the optical parameters of the water, is simulated with Geant4 \cite{Hu:2023ife}. By convolving the two PDFs we obtained a realistic photon arrival time distribution model, which can be used to fit the photon arrival time distribution data collected by T-REX in deep sea and get the optical parameters from the optimal fitting results.

\section{Performance of the PMT System} \label{sec:performance_pathfinder}

In the sea experiment, the T-REX apparatus was deployed at a depth of $3420~\mathrm{m}$. The whole apparatus was suspended for about 2 hours to conduct \textit{in-situ} measurements, during which each PMT detected more than $10^7$ photons for each wavelength. During the data collection, all PMTs in the two LRMs and the pulsing LEDs in the LEM were synchronically triggered by the WR system at a frequency of $10~\mathrm{kHz}$. 

Since the distances between LEDs and PMTs are fixed, the peak of the LED light signals received by each PMT always has a predicted time delay after the LED light emission. By stacking a large number of trigger windows for each PMT, a prominent peak of LED light signals can be found. The noise, either the PMT dark noise or the ambient light background, occurs randomly in time. So these noise result in a flat background after stacking a large number of trigger windows. The photon arrival time distribution shown in Figure \ref{fig:arr_time} is obtained by subtracting the flat background. The water optical properties, including absorption and scattering length can be extracted from the photon arrival time distributions of near and far PMTs, as presented in Method section in \cite{Ye:2022vbk}

\begin{figure}[htb]
    \centering
    \includegraphics[width=0.98\textwidth]{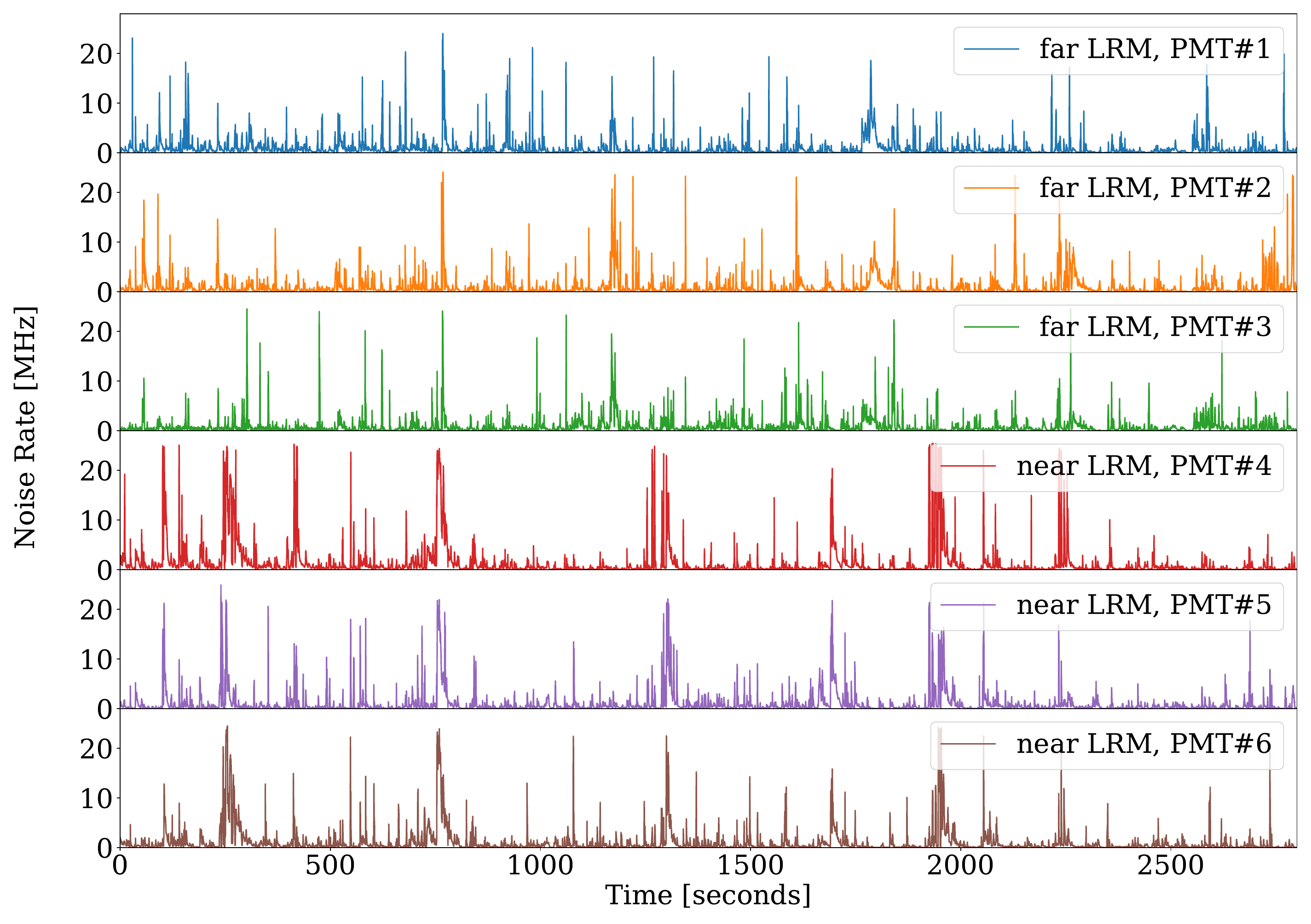}
    \caption{The noise rates of the six PMTs in the two LRMs during data collection with T-REX.}
    \label{fig:noise_rate}  
\end{figure}

The noise rates of the six PMTs as a function of operation time in the sea experiment are shown in Figure \ref{fig:noise_rate}. It can be seen that the noise rate of the PMTs in the same LRM are strongly correlated in time. This correlated noise rate is likely caused by the strong vibration of the T-REX apparatus due to suspending from the ship. If a high noise rate occurs on a PMT, the data for all PMTs in that trigger window is removed. Hence only the data in low noise period is selected for the stacking analysis to increase the signal-to-noise ratio.

A near-far relative measurement was performed to determine the effective attenuation length of light in deep-sea water with T-REX data. The expected number of PEs detected by a PMT at distance $d$ is given by: 
\begin{equation}
    N_{\text{tot}} = N_{\text{e}} \cdot \frac{A_{\text{r}}}{2\pi d^{2}} \cdot e^{-\frac{d}{\lambda_{\text{eff, att}}}} \cdot \eta_{\text{r}},
\end{equation}
where $N_{\text{e}}$ and $N_{\text{tot}}$ are the numbers of emitted and detected photon, $A_{\text{r}}$ and $\eta_{\text{r}}$ are the area and the detection efficiency of the receiver PMT. Therefore, the effective attenuation length of the water, $\lambda_{\text{eff, att}}$, can be derived by comparing the total number of photons detected by two PMTs at different distances:
\begin{equation}
   \lambda_{\text{eff, att}} = (d_\text{n} - d_\text{f}) /\ln{(\frac{N_{\text{tot,n}}}{N_{\text{tot,f}}} \frac{N_{\text{e,f}}}{N_{\text{e,n}}} \frac{d_\text{n}^2}{d_\text{f}^2} \frac{\eta_\text{r,n}}{\eta_\text{r,f}})},
\end{equation}

where the indices $n$ and $f$ denote the variables at near and far distances of $d_\mathrm{n} = 21.56~\text{m}$ and $d_\mathrm{f}=41.62~\text{m}$, respectively.
$N_{\text{e,n}}/N_{\text{e,f}}$ is the LED brightness ratio, and $\eta_\text{r,n} / \eta_\text{r,f}$ is the relative PDE of the two PMTs in the near and far LRMs. These two factors are calibrated in the lab, as presented in Section \ref{sec:calibration}. 

The absorption and scattering lengths of light in water are extracted by performing chi-square fitting of the measured photon arrival time distribution with the model constructed by convolution of the calibrated distribution in air shown in Figure \ref{fig:arr_time} and simulated photon propagation PDF \cite{Hu:2023ife}. Details of the analysis can be found in the paper \cite{Ye:2022vbk}.

\section{Conclusion and Outlook}
\label{sec:conclusion}

In this paper, we introduce the PMT detection system designed for the TRIDENT pathfinder experiment. The system consists of six 3-inch PMTs in the two LRMs at different distances from the LEM to conduct a near-far relative measurement of the deep-sea water optical properties at the pre-selected site for TRIDENT. Testing and calibration of the PMTs quantitatively ensure the success of the measurement. By counting the number of photons and the arrival time distribution measured by the PMTs, the attenuation length, absorption length and scattering length of light in the deep-sea water are measured for three wavelengths. The successful measurement of the sea water optical properties serves as a basis for future neutrino telescope design as presented in \cite{Ye:2022vbk}. The pathfinder experiment also verified that the PMT detection system works well in the deep sea.

PMTs are fundamental detection units and are critical for the success of a neutrino telescope. The use of multiple 3-inch PMTs in a digital optical module and the techniques of assembling the module in T-REX set a solid foundation for the following R\&D phase of TRIDENT. The design of T-REX, using pulsing LEDs and PMTs, can be a potential solution for calibrating water optical properties for TRIDENT in real time \cite{IceCube:2013llx}.

For TRIDENT, we aim to use PMTs with higher QE, lower DCR, smaller TTS and better SPE resolution to enhance the overall performance of the neutrino telescope. The length of the PMT will also be optimized to integrate more PMTs in a DOM. Furthermore, hDOMs with multiple 3-inch PMTs and SiPM arrays are under development to provide improvement in TRIDENT's angular resolution. The multiple photon sensors in each hDOM also allow implementing local coincidence trigger algorithm to reduce backgrounds. A larger target volume, lower background rate, and better pointing resolution will make TRIDENT a next-generation neutrino telescope with significantly higher sensitivity for the search of high energy astrophysical neutrinos.

\section*{Acknowledgement}

We thank Jun Guo and Xin Xiang for their help to improve this paper. This work is supported by the Ministry of Science and Technology of China [No. 2022YFA1605500]; Office of Science and Technology, Shanghai Municipal Government [No. 22JC1410100]; and Shanghai Jiao Tong University under the Double First Class startup fund and the Foresight grants [No. 21X010202013] and [No. 21X010200816].


\providecommand{\href}[2]{#2}\begingroup\raggedright\endgroup

\end{document}